# Coupled Real- and Momentum-Space Topology in Symmetry-Locked Bilayer Altermagnet

Shuo Zhang,[1] Zijie Fu,[2] Lixiu Guan,[3] Yirui Du,[1] Linyang Li,[1,3,*] and Junguang Tao[1,4,†]

[1]*School of Materials Science and Engineering, Hebei University of Technology, Tianjin 300132, China*

[2]*Arizona College of Technology at Hebei University of Technology, Tianjin 300401, China*

[3]*School of Science, Hebei University of Technology, Tianjin 300401, China*

[4]*Hebei Engineering Laboratory of Photoelectronic Functional Crystals, Hebei University of Technology, Tianjin 300132, China*

**Abstract:** Integrating real-space topological spin textures with momentum-space topological electronic states within a single altermagnetic system has remained a persistent challenge. Here, we introduce a symmetry-locked bilayer altermagnet that concurrently hosts $d$-wave altermagnetism, momentum-space topology, and stable antiskyrmions. In momentum-space, it enables strain-triggered transitions to an antiferromagnetic Weyl semimetal phase, where the Néel vector acts as a switch for spin-layer-polarized quantum anomalous Hall and Weyl states, alongside coupled topological states and valley polarization effects. In real-space, the formation of interlayer co-directional and locked in-plane Dzyaloshinskii-Moriya interactions facilitates the creation of coupled antiskyrmion pairs with compensated topological charges. This locking symmetry fully cancels the transverse Magnus forces, resulting in current-driven, strictly longitudinal motion of antiskyrmions without any Hall-like deflection. Our work establishes a robust platform for dual-space topological magnetism and offers a definitive solution to the vanishing Hall angle in charge-neutral antiskyrmions, opening pathways toward high-density, low-power topological spintronics.

*Contact author: linyang.li@hebut.edu.cn

†Contact author: jgtao@hebut.edu.cn

Condensed matter physics is pursuing the integration of two distinct topological paradigms: real-space topology emerging from magnetic quasiparticles (e.g., skyrmions, bimerons, and antiskyrmions) for spintronics[1-5], and momentum-space topology of electronic bands with nontrivial spin-orbital coupling (SOC) gaps, responsible for robust quantum transport phenomena like quantum spin/anomalous Hall (QSH/QAH) effects[6-12]. In two-dimensional (2D) ferromagnetic (FM) systems, the magnetic-order orientation acts as a tunable degree of freedom that modifies symmetry constraints, allowing gapless points to open up by SOC and facilitating a transition from a gapless Weyl semimetal to a gapped Chern insulator[13,14]. These manifestations of topology have typically been realized in distinct material families. The recent emergence of altermagnetism[15-19], which combines time-reversal symmetry breaking with high crystal symmetry, presents a promising platform to bridge this divide. The central, yet unrealized, challenge is to simultaneously host and couple nontrivial topological order in both spaces within a single 2D altermagnet. Achieving such "topological duality" would establish a unified material platform where protected spin textures and topological electronic transport coexist and interact, paving the way for emergent magnetoelectric phenomena and advanced quantum technologies.

The skyrmion/antiskyrmion Hall effect (SkHE) driven by Magnus force deflects its motion perpendicular to the current direction, inevitably annihilated at edges, destroying the information carrier and precluding reliable device operations[20,21]. The compensated magnetic order in altermagnetic platform allows for theoretical existence of two sets of real-space topological spin textures with opposite topological

charges,—forming a "topologically neutral" pair—resulting in a nearly zero SkHE[22]. This locked-directional-motion is the holy grail for information encoding and transmission using topological spin textures, promising ultra-dense, low-power, and ultra-reliable memory and logic devices. Realizing such compensation typically requires the formation of a co-directionally locked Dzyaloshinskii–Moriya interaction (DMI), a condition that depends on specific Moriya symmetry constraints and therefore remains nontrivial to achieve.

The design of 2D materials for integrated real- and momentum-space topology leverages intrinsic structural asymmetries to generate DMI and Berry curvature[23,24]. By engineering spin-space group symmetries, compensated magnetic order and $k$-dependent spin splitting can be encoded. Controlled symmetry breaking *via* stacking or strain offers routes to concurrently induce DMI and topological bands. However, realizing such system require a delicate balance of exchange, anisotropy, and DMI to stabilize, for example, skyrmions within an altermagnetic host and to lock them into compensated pairs.

In this Letter, we successfully achieved locked linear motion of topological spin textures in a bilayer A-type antiferromagnetic (A-AFM) system that simultaneously hosts altermagnetism (~550 meV splitting), topological spin structures in both real- and $k$-space. Strain can drive the system into an altermagnetic QSH insulator phase for out-of-plane magnetization, where spin-layer-polarized properties are explicitly manifested. Upon applying in-plane magnetization, the Néel vector should break/preserve the mirror symmetry under SOC. This yields a nontrivial bandgap in one spin channel and a gapless semimetal state in the other, corresponding to spin-layer-polarized QSH and Weyl phases, respectively. Furthermore, the symmetry-inherent co-directional DMI enables complete cancellation of Hall response in antiskyrmion pairs, establishing a new paradigm for zero-field, topologically protected spintronics.

To clarify the design principle, we commence with a symmetry-based analysis aimed at achieving three concurrent objectives: multiple topological phases, altermagnetism, and elimination of SkHE. We adopt $D_{2d}$ symmetry as a non-exclusive yet robust framework, as it is compatible with $d$-wave altermagnetic responses, anisotropic DMI, and two orthogonal spin-degenerate nodal planes[25-27]. The A-AFM configuration enables complete compensation of transverse transport while preserving topological stability of magnetic structure[28]. In bilayer system, when the spins in top layer (TL) and bottom layer (BL) are antiparallel but share identical chirality, the total topological charge satisfies $Q_{tot} = 0$. Therefore, both the transverse velocity and associated topological Hall response vanish[29] due to the unidirectional locking of the layer-resolved DMI.

To achieve this, as illustrated in Fig. 1(a), the bilayer structure is constructed in an operator-based manner: TL is symmetry-generated from BL. We introduce a stacking operator $\widehat{P} = P\tau_z\tau_O$, where $P$ denotes a point-group operation, $\tau_z$ represents an out-of-plane translation, and $\tau_O$ is in-plane fractional shift. The $P$ can be regarded as a class of 2D symmetry operations that preserve layer orientation, $Q^+ = \{E, C_z(\theta), \mathcal{M}_\beta\}$, as opposed to the orientation-reversing class $Q^- = \{C_{2\alpha}, \mathcal{M}_z, I, S_n^z\}$[30]. Accordingly, TL can be generated from BL *via* a Seitz-type mapping $\boldsymbol{r} \mapsto P\boldsymbol{r} + \boldsymbol{\tau}$, which formally establishes a one-to-one correspondence between the local coordinates $(x_{BL}, y_{BL})$, $(x_{TL}, y_{TL})$, and the global coordinates $(x, y)$. In the present system, the specific point-group operation is given by the composite operator $P = \mathcal{M}_z C_{4z}$, followed by an interlayer translation $\tau_z = 2t_z$ to spatially separate the two layers, with no relative in-plane displacement ($\tau_O = 0$). In this sense, the formation of the bilayer system is essentially the result of a constrained symmetry mapping combined with prescribed interlayer and intralayer translations.

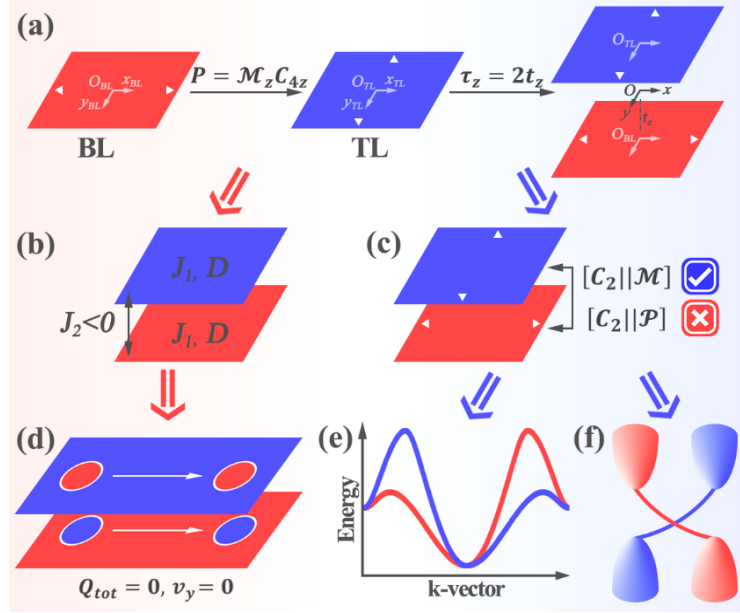

**FIG. 1.** Symmetry-locked design of co-directional DMI and altermagnetism. (a) Bilayer A-AFM structure constructed from monolayer *via* stacking operator $\widehat{P} = P\tau_z\tau_O$. The white triangles highlight the atoms in unit cell that transform under the corresponding symmetry operations. (b) Schematic illustrating co-directional locking of intralayer DMI in TL and BL. (c) Diagram of symmetry relations governing the bilayer system. (d) Illustration of suppression of SkHE. (e) Characteristic spin-split band structure of altermagnetism. (f) Schematic representation of spin-layer-polarized QSH insulator.

Noncollinear spin textures typically originate from DMI, with the Hamiltonian given by $H = -\sum_{i,j} \boldsymbol{D}_{ij} \cdot (\boldsymbol{S}_i \times \boldsymbol{S}_j)$. In monolayer and bulk systems, breaking inversion symmetry is typically required to realize a finite DMI. However, in bilayer systems that preserve global inversion symmetry, a nonzero net DMI can still exist within each individual layer. This seemingly paradoxical scenario can be resolved by the transformation behavior of $\boldsymbol{D}$ vector under symmetry operations. We consider a symmetry operator $\hat{A} = \{A|t\}$ in space group $G$ acting on $\boldsymbol{D}$ vector, where $A \in SO(3)$ represents the rotational part and $t$ the translational part[29]. Under this operator, the transformed DMI vector in TL ($\boldsymbol{D}^{TL}$) satisfies the relation $\boldsymbol{D}^{TL} = det[A]A\boldsymbol{D}^{BL}$. As shown in Supplemental Material Note II, for a global inversion operation ($A = -I$), we obtain: $\boldsymbol{D}^{TL} = -\boldsymbol{D}^{BL} = (-D_x, -D_y, -D_z)^T$. In contrast, under a mirror symmetry (e.g., mirror plane perpendicular to the $z$-axis), $\boldsymbol{D}^{TL} = (-D_x, -D_y, D_z)^T$. Therefore, in A-AFM bilayer that preserves either global inversion or mirror symmetry, $\boldsymbol{D}$ vectors in TL and BL must have opposite signs. This implies that a globally inversion-symmetric A-AFM structure inherently cannot achieve cancellation of the Hall deflection, as the required co-directional locking of layer-resolved DMI is forbidden.

To overcome this constraint, we introduce the composite symmetry operation in Fig. 1(a), which simultaneously breaks both global inversion and mirror symmetries. Under this operation, the transformed DMI vector follows: $\boldsymbol{D}^{TL} = (D_x, -D_y, D_z)^T$, corresponding to an orthogonal exchange of in-plane components. This transformation is inherently compatible with anisotropic DMI that stabilizes antiskyrmions, yet it is incompatible with isotropic DMI required for Néel-type or Bloch-type skyrmions.

Consequently, the antiskyrmion $\boldsymbol{D}$ vectors satisfy: $D_x^{TL} = D_y^{BL}$ and $D_y^{TL} = D_x^{BL}$. Thus, the stacking operator $\hat{P}$ realizes the co-directional locking of DMI vectors. In Fig. 1(b), the resulting structure exhibits interlayer antiferromagnetic alignment and intralayer ferromagnetic coupling, while DMI in both layers is locked to the same chirality. In such system, where spin interactions support a chiral spin topology and can host antiskyrmions, the Magnus forces is mutually compensated [Fig. 1(d)], effectively suppressing the SkHE.

From the perspective of band topology and magnetic space-group constraints, the two layers are established by $\hat{P}$, which cannot restore global $\mathcal{PT}$ symmetry. Consequently, A-AFM bilayers serve as a versatile stacking platform to induce altermagnetism. As illustrated in Fig. 1(c), $\hat{P}$ breaks $[C_2||\mathcal{P}]$ symmetry while preserving $[C_2||\mathcal{M}]$ constraint. This favors the emergence of altermagnetism, governed by the condition $E(s, \boldsymbol{k}) = E(-s, \mathcal{M}\boldsymbol{k})$, whereas the conventional $\mathcal{PT}$-enforced degeneracy $E(s, \boldsymbol{k}) = E(-s, -\boldsymbol{k})$ is lifted. As a result, spin splitting is no longer trivially compensated between $\boldsymbol{k}$ and $-\boldsymbol{k}$. Instead, it forms anisotropic paired distributions between $\boldsymbol{k}$ points related by rotation or mirror operations. As shown in Fig. 1(e), the system exhibits spin-resolved band structures across the Brillouin zone.

To validate this, we select Janus VSSe as a representative prototype. Monolayer Janus VSSe (see Fig. S1 in the Supplemental Material [31]) possesses C$_{2v}$ symmetry and belongs to the $P\bar{4}m2$ (No. 115) space group. Two distinct interlayer configurations are considered to construct the bilayer A-AFM system. In the first one, TL is generated from BL *via* $\mathcal{M}$ operation [Fig. 2(a)]. According to Moriya's rules[32], the resulting DMI vectors satisfy: $\boldsymbol{D}^{TL} = (-D_x, -D_y, D_z)^T$ and $\boldsymbol{D}^{BL} = $ $(D_x, D_y, D_z)^T$, indicating their opposite signs of in-plane components. Therefore, effective Hall-effect suppression cannot be achieved. Simultaneously, this stacking preserves both $[C_2||\mathcal{P}]$ and $[C_2||\mathcal{M}]$, enforcing $E(s, \mathbf{k}) = E(-s, \mathbf{k})$. As shown in Fig. 2(b), the spin degeneracy is fully restored, and the system behaves as a conventional antiferromagnet rather than an altermagnet. In the second configuration, TL is generated from BL via $P = \mathcal{M}_z C_{4z}$ operation [Fig. 2(c)]. Thus, the in-plane components of DMI vectors in TL and BL remain co-aligned, and symmetry analysis further gives $D_x = -D_y = D$. Consequently, a unidirectional locking of layer-resolved DMI is achieved, establishing essential prerequisite for suppressing SkHE in a bilayer A-AFM system.

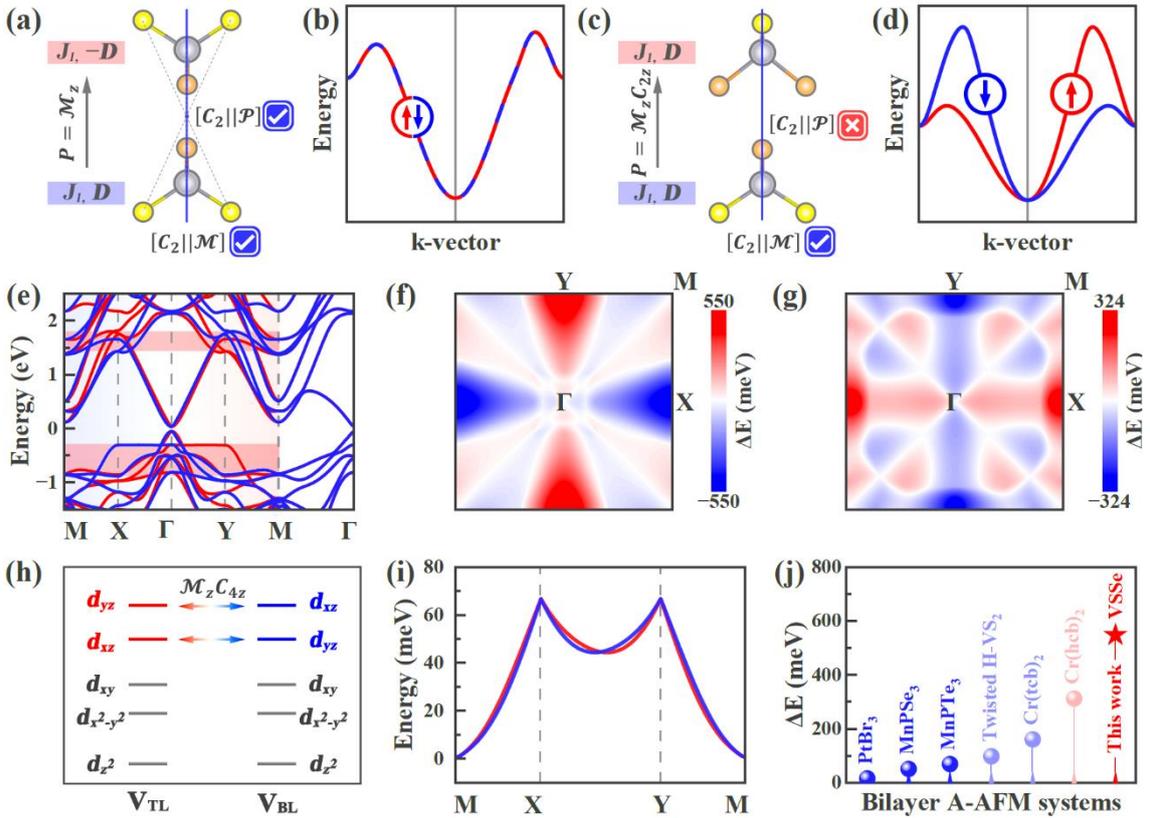

**FIG. 2.** Altermagnetic characteristics of bilayer VSSe system. (a) Bilayer stacking configuration preserving $[C_2||\mathcal{P}]$ symmetry, where intralayer DMI is oppositely locked. (b) Corresponding spin-degenerate band structure. (c) Bilayer stacking

configuration with broken $[C_2||\mathcal{P}]$ symmetry, leading to co-directionally locked intralayer DMI. (d) Spin-split band structure. (e) Electronic band structure of (c), where red (blue) denotes spin-up (spin-down). Spin splitting of (f) VB and (g) CB across entire Brillouin zone. (h) Schematic crystal-field diagram. (i) Magnon spectrum exhibiting altermagnetic splitting features. (j) Comparison of band-splitting magnitudes in bilayer A-AFM system.

Given that weak van der Waals interlayer coupling permits possible twisting or lateral sliding, multiple stacking symmetries can arise. We therefore systematically explored the stacking configurations in this scenario (Fig. S2), with the most stable one marked by a yellow pentagram. Phonon dispersion calculations confirm the structural stability for both monolayer and bilayer A-AFM configurations (Fig. S3). Notably, the Néel temperature is well above room temperature (see Fig. S4). Differential spin-density analysis reveals that the interfacial spin charge undergoes a spatially symmetric redistribution, preserving $\mathcal{P}$-symmetry in real-space (Fig. S5). Therefore, altermagnetism primarily stems from spin-space symmetry breaking, rather than charge asymmetry.

The electronic bands near the Fermi level ($E_F$) are displayed in Figure 2(e), where a clear spin-splitting is observed. Two representative regions are highlighted by red boxes: the bands remain spin-degenerate at $\Gamma$ point, whereas pronounced splitting emerges near $X$ and $Y$ points along $M$–$X$–$\Gamma$–$Y$–$M$ path. The spin-splitting across the entire Brillouin zone is presented in Fig. 2(f, g). The results reveal substantial splitting in both valence and conduction bands, with maxima located at $X/Y$ points reaching

~550/324 meV. The crystal field splits *d* orbitals into five non-degenerate levels [Fig. 2(h)]. Upon $\mathcal{M}_z C_{4z}$ operation, $d_{xy}$, $d_{x^2-y^2}$ and $d_{z^2}$ orbitals remain invariant, whereas $d_{xz}$ and $d_{yz}$ are interchanged. The resultant orbital exchange ultimately drives the altermagnetic behavior. The layer- and spin-resolved projected band structures shown in Fig. S6 further demonstrate the splitting of relevant bands, corroborating the symmetry analysis above. Due to symmetry breaking, the magnon spectrum also exhibits splitting features analogous to those in the electronic bands. In Fig. 2(i), the maximum magnon splitting is ~3.2 cm$^{-1}$, which is experimentally accessible, although smaller than that in bulk MnTe and RuO$_2$[33,34]. As shown in Figs. S7-8, under −4% to 4% strain, the system exhibits a semiconductor-semimetal transition while maintaining altermagnetic characteristics, with maximum spin splittings reaching ~578 meV (valence, 2% tension) and ~480 meV (conduction, −4% compression)—values exceeding those of most reported bilayers [Fig. 2(j)] [15,35-38], highlighting the potential of strain for tunable spintronic modulation.

Considering SOC effect, when the Néel vector $\hat{\boldsymbol{n}} = \hat{\boldsymbol{m}}_{BL} - \hat{\boldsymbol{m}}_{TL}$ is along [001] direction, the band gap (45.1 meV) of VSSe bilayer remains similar to that without SOC, and altermagnetism is preserved [Fig. 3(a)]. Under –2% strain, the system without SOC exhibits four spin-polarized gapless Weyl points along *Γ*-*X*(-*Y*), protected by $\mathcal{M}_y(\mathcal{M}_x)$ mirror symmetries [Fig. S7(b)]. Since $\hat{\boldsymbol{n}}$ is an axis vector, [001] orientation breaks both $\mathcal{M}_y$ and $\mathcal{M}_x$, opening a nontrivial SOC bandgap of 37.8 meV at each Weyl point. Owing to the spin-layer-polarized nature of altermagnetism, each gapped Weyl point in TL (spin-down, blue) contributes a quantized Berry phase of −π, accumulating to a total of −2π. Conversely, in BL (spin-up, red), each contributes +π, summing to 2π. This results in a spin-layer-polarized QSH effect with Chern numbers $C = -1$ in TL and $C = 1$ in BL [Figs. 3(g) and 3(j)]. The above physical process can be confirmed by calculating the spin Hall conductivity (SHC) using Kubo-Greenwood formula:[39-41]

$$\sigma_{xy}^{S_z}(\omega) = \hbar \int_{BZ} \frac{d^3k}{(2\pi)^3} \sum_n \sum_{m \neq n} 2 f_n(\mathbf{k}) \frac{Im \langle \psi_{n\mathbf{k}} | \hat{j}_x^{S_z} | \psi_{m\mathbf{k}} \rangle \langle \psi_{m\mathbf{k}} | -e\hat{v}_y | \psi_{n\mathbf{k}} \rangle}{(E_{n\mathbf{k}} - E_{m\mathbf{k}})^2 - (\hbar\omega + i\eta)^2} \quad (1)$$

where $\hat{j}_x^{S_z}$ represents the spin current operator projected along z-direction. Under the direct current limit ($\omega, \eta \rightarrow 0$), SHC quantizes to 2/0 (in units of e/4π) under –2%/0% strain, accompanied by a reversal of Berry curvature [Figs. 3(c) and 3(f)]. The SHC can be defined as $\sum_{l=1}^{2}(-1)^l C_l$, where $C_l$ is the layer Chern number.

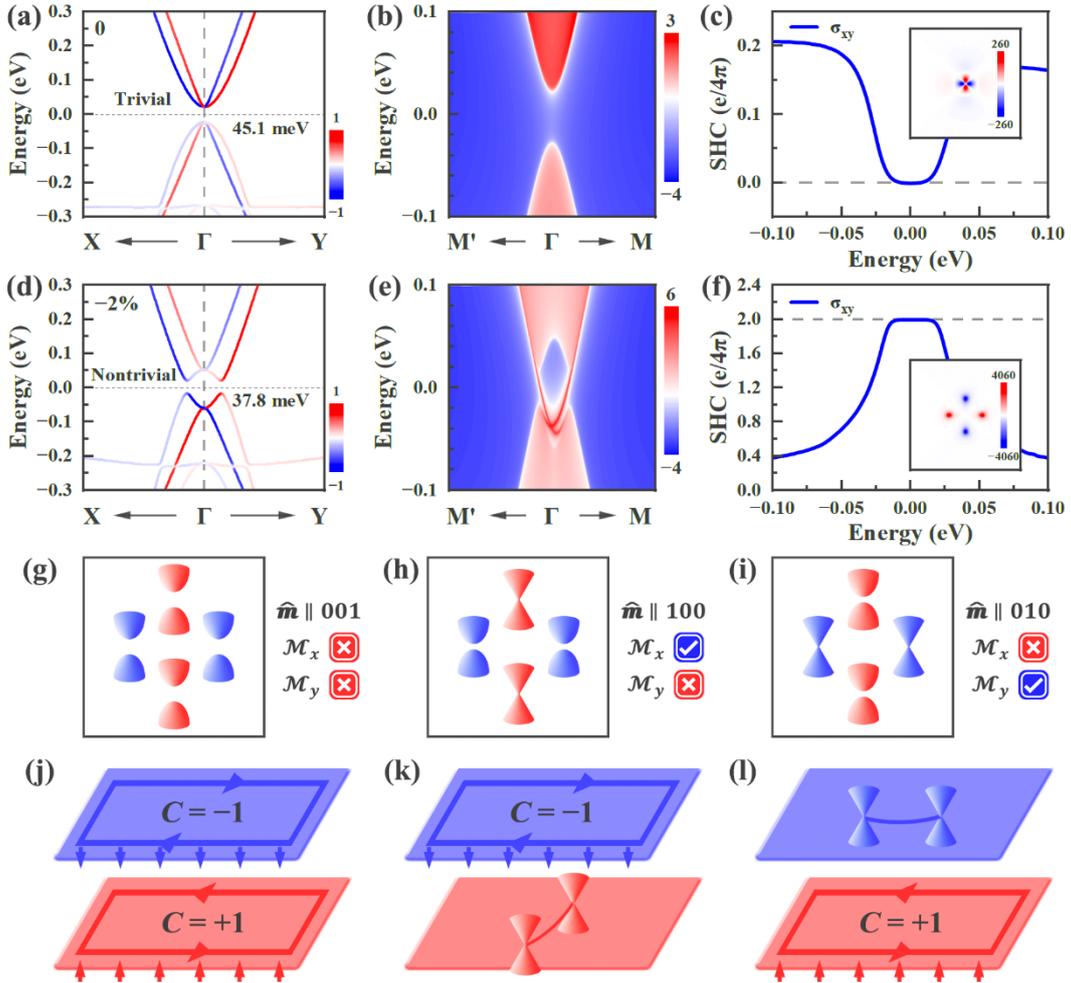

**FIG. 3.** Strain-driven topological and transport transitions. (a) Spin-resolved band

structure, (b) edge states, and (c) SHC and Berry curvature in the absence of a force field. (d) Spin-resolved band structure, (e) edge states, and (f) SHC and Berry curvature under −2% strain. (g-i) Schematic distribution of spin-resolved Weyl cones in Brillouin zone with the Néel vector aligned along $z$, $x$, and $y$, respectively. Conical and cap-shaped features denote occupied and unoccupied states. (j-l) Schematic illustration of spin-layer-polarized QSH state for the corresponding Néel-vector orientations.

The strain-driven altermagnetic QSH effect is further corroborated by edge-state calculations, which reveal two crossing edge states connecting the valence and condition bands [Figs. 3(b) and 3(e)]. Here, the system symmetry plays a crucial role. When $\hat{\bm{n}}$ is oriented along [100] and [010] directions, the spin-layer-polarized topological properties emerge. For $\hat{\bm{n}} = [100]$, the two magnetic moments $\hat{\bm{m}}_{TL}$ and $\hat{\bm{m}}_{BL}$ both preserve $\mathcal{M}_x$ symmetry, but break $\mathcal{M}_y$ symmetry. Consequently, only two gapless Weyl points appear along high-symmetry path $\Gamma$–$Y$ (parallel to $\mathcal{M}_x$), while nontrivial bandgap (13.8 meV) opens along $\Gamma$-$X$ due to the broken $\mathcal{M}_y$ symmetry. Thus, TL behaves as a Chern insulator with $C = -1$ whereas BL remains a Weyl semimetal [Figs. 3(h) and 3(k)]. For $\hat{\bm{n}} = [010]$, the roles reverse: BL becomes a Chern insulator with $C = +1$ while TL is a Weyl semimetal, as shown in Fig. S10. Under an applied strain of −4%, the system exhibits topological states analogous to those at −2% strain, with the nontrivial bandgap further increasing to 62.4 meV for [001] direction [Fig. S9(a)].

In above stain-driven topological phase transition, we also noticed the change of Berry curvatures, which is calculated by[42,43]

$$\Omega(\boldsymbol{k}) = -\sum_n \sum_{n \neq m} 2f_n(\boldsymbol{k}) \frac{Im\langle\psi_{n\boldsymbol{k}}|v_x|\psi_{m\boldsymbol{k}}\rangle\langle\psi_{m\boldsymbol{k}}|v_y|\psi_{n\boldsymbol{k}}\rangle}{(E_{n\boldsymbol{k}} - E_{m\boldsymbol{k}})^2} \quad (2)$$

where $f_n$ is the Fermi-Dirac distribution function, and $v_x(v_y)$ is the velocity operator along $x(y)$-direction. $\psi_{nk}$ represents the Bloch wave function with eigenvalue $E_n$. Along a given high-symmetry path, a reversal in the sign of Berry curvature not only signals a topological phase transition but also modifies the transport properties. Here, we focus on $\hat{n} = [001]$ for interpretation. In the absence of strain, the Berry curvature along XX′ is negative, corresponding to the spin-down channel of TL, while along YY′ it is positive, corresponding to the spin-up channel of BL. In other words, the sign of the Berry curvature is locked to both spin and layer indices. This nonzero Berry curvature induces an anomalous velocity for Bloch electrons in the valleys under an applied electric-field $\boldsymbol{E}$, following the relation $\boldsymbol{v} \sim \boldsymbol{E} \times \Omega(\boldsymbol{k})$[44]. The system thereby exhibits a spin-valley Hall effect, as illustrated in Fig. S11. Upon applying a force field (−2% and −4%), the VSSe bilayer itself hosts a large Berry curvature (~4060 and ~1890 Å$^2$), with opposite signs emerging in the two spin channels at the four valleys (X, X′, Y and Y′). This leads to a pronounced spin-valley Hall effect, enabling a net spin current that is accessible in experimental detection[45]. Notably, the spatial region in which spin current accumulates is completely reversed compared to the unstrained case. At −4% strain, the two spin channels acquire anomalous velocities in opposite directions, which are further locked to the respective layers, giving rise to a spin-layer-polarized valley Hall effect. By tuning the direction of in-plane $\boldsymbol{E}$, the spin current direction can be correspondingly modulated. In short, the strained altermagnetic VSSe bilayer achieves the coexistence of a QSH state and valley polarization effects, demonstrating promising potential for the new type of spin electronics devices.

A 2D atomic spin Hamiltonian model is constructed as:

$$H = H_{intra}^{TL} + H_{intra}^{BL} + H_{inter} \quad (3)$$

$$H_{intra}^{TL(BL)} = -J_1 \sum_{\langle i,j \rangle} \mathbf{S}_i \cdot \mathbf{S}_j - K \sum_i (S_i^z)^2 - \sum_{\langle i,j \rangle} \mathbf{D}_{ij} \cdot (\mathbf{S}_i \times \mathbf{S}_j)$$
$$- \sum_i \mu_i \mathbf{B} \cdot \mathbf{S}_i \quad (4)$$

$$H_{inter} = -J_2 \sum_{\langle i \in TL, j \in BL \rangle} \mathbf{S}_i \cdot \mathbf{S}_j \quad (5)$$

All magnetic parameters are summarized in Table I, with computational details provided in Supplemental Material Note V. As shown in Fig. 4(a), TL and BL are antiferromagnetically coupled and each host stable topological magnetic textures. In the compressive strain range (−4% to 0%), the regularly arranged antiskyrmions in both layers form an ordered antiskyrmion lattice. Under tensile strain (2% to 4%), the system progressively transitions into a mixed phase, where labyrinthine domains coexist with antiskyrmions. The $Q$ of two layers exhibits mirror-symmetric evolution with strain [Fig. S13(a)]: it remains nearly constant under compression but gradually decreases under tension. Correspondingly, the antiskyrmion radius $R$ [Fig. S13(b)] reaches maximum at zero strain and contracts slightly under tensile strain. With increasing strain, the DMI energy decreases gradually while the magnetic anisotropy energy rises slowly [Fig. S13(c)]. The competition between these two energy terms collectively governs the stability and size evolution of the topological texture. Overall, strain enable precise regulation of topological magnetic properties.

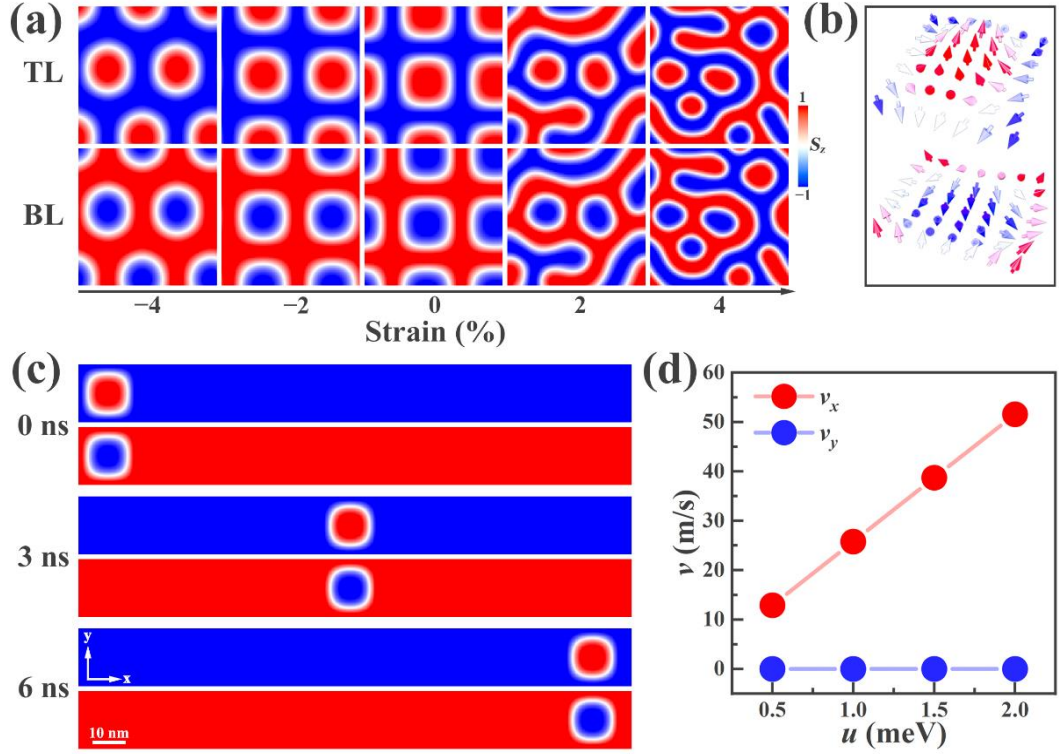

**FIG. 4.** Suppression of SkHE. (a) Stable topological spin textures formed in bilayer VSSe under different strain conditions. (b) Schematic spin configuration of interlayer antiskyrmion pair. (c) Current-driven motion snapshots of the antiskyrmion pair at 0, 3, and 6 ns. (d) Dependence of the motion velocity on $u$.

TABLE I. Magnetic parameters of bilayer VSSe under different strain conditions

|  | $J_1$ (meV) | $J_2$ (meV) | $K$ (μeV) | $D$ (meV) | a (Å) |
|---|---|---|---|---|---|
| -4% | 37.93 | -4.96 | 68 | 3.45 | 3.605 |
| -2% | 35.92 | -5.71 | 33 | 3.88 | 3.681 |
| 0 | 34.26 | -6.36 | 15 | 4.87 | 3.756 |
| 2% | 32.49 | -6.94 | -11 | 6.01 | 3.831 |
| 4% | 30.31 | -7.46 | -38 | 7.17 | 3.906 |

Figure 4(b) presents a schematic spin-vector configuration of a representative antiskyrmion. As designed, symmetry-locking results in net zero overall topological charge ($Q_{tot} = 0$). When driven by electric current, the antiskyrmion dynamics are governed by the Landau-Lifshitz-Gilbert (LLG) equation[46,47]:

$$\boldsymbol{G} \times \boldsymbol{v} - \alpha \boldsymbol{\mathcal{D}} \cdot \boldsymbol{v} + 4\pi \boldsymbol{\mathcal{B}} \cdot \boldsymbol{j} = 0 \qquad (5)$$

The first term, Magnus force, is directly related to SkHE and gyrocoupling vector $\boldsymbol{G} = (0, 0, 4\pi Q)$ is determined by $Q$. The velocity $\boldsymbol{v} = (v_x, v_y)$ describes the motion of the antiskyrmion. Applying an in-plane current along $-y$ direction via spin-transfer torque ($u$ = 1 μeV) and solving Eq. (4), we obtain $v_x = [\alpha \mathcal{D}_{yy}/(Q^2 + \alpha^2 \mathcal{D}_{xx}\mathcal{D}_{yy})]\mathcal{B}j$, and $v_y = [Q/(Q^2 + \alpha^2 \mathcal{D}_{xx}\mathcal{D}_{yy})]\mathcal{B}j$. For A-AFM bilayer system with $Q_{tot} = 0$, once an antiskyrmion is formed in BL, TL simultaneously hosts an antiskyrmion carrying the opposite topological charge ($Q_{TL} = -Q_{BL}$)[48]. This ensures synchronized motion along $x$-direction ($v_x^{TL} = v_x^{BL}$), while $v_y$-components are antiparallel ($v_y^{TL} = -v_y^{BL}$). Consequently, the transverse velocity contributions cancel exactly, fundamentally suppressing the SkHE. Figure 4(c) illustrates the current-driven dynamical evolution of the isolated antiskyrmion pair. It exhibits pronounced suppression of SkHE and moves stably along $+x$-direction, achieving a longitudinal velocity of $v_x$ = 25.80 m/s. Importantly, a linear velocity–current relation is established in Fig. 4(d) to enable controllable driving and device-level regulation of the topological texture motion.

It needs to be emphasized that even when opposite layer charges yield $Q_{tot} = 0$ and cancel macroscopic Hall signals in bilayer systems, layer-resolved topological responses or spin- and orbital-related transport features may still persist[49], distinct from the topological layer Hall effect[50]. Within this context, the two interlayer antiskyrmion pair, $(Q_{TL}, Q_{BL}) = (+1, -1)$ and $(-1, +1)$, represent internally inverted yet macroscopically equivalent compensated states. Although their overall dynamics are identical, layer exchange reverses the topological distribution, enabling topology-based encoding *via* layer degrees of freedom.

The analysis above demonstrates that in an A-AFM bilayer, effective suppression of the Hall effect requires the generation of nonreciprocal velocity components along *y*-direction between TL and BL. This is achieved by selectively breaking interlayer $[C_2||\mathcal{P}]$ symmetry while preserving $[C_2||\mathcal{M}]$, maintaining zero net magnetization to avoid stray magnetic fields. Our symmetry analysis confirms that these conditions can be met simultaneously through a rationally designed interlayer stacking operation.

In summary, we demonstrate that asymmetric stacking in A-AFM VSSe bilayer with broken $[C_2||\mathcal{P}]$ symmetry induces *d*-wave altermagnetism characterized by pronounced spin splitting of ~550 meV. Notably, this altermagnetic order exhibits both spin-layer polarization and momentum-space topology. Strain further triggers a transition to an antiferromagnetic Weyl semimetal phase, wherein the Néel vector acts as a switch for spin-layer-polarized quantum anomalous Hall and Weyl states. The symmetry-locked DMI not only stabilizes layer-resolved chiral spin textures but also completely cancels the transverse dynamics of interlayer antiskyrmion pairs, eliminating the Hall offset and enabling a maximum velocity of ~25.8 m/s. This work establishes a clear paradigm for designing 2D magnetic systems that host dual-space topological magnetism and altermagnetic order, paving the way for future spintronic applications.

## Acknowledgements

This work was supported by Natural Science Foundation of Hebei Province (A2025202003 and E2024202044).

# References

[1] C. Xu, P. Chen, H. Tan, Y. Yang, H. Xiang, and L. Bellaiche, Electric-Field Switching of Magnetic Topological Charge in Type-I Multiferroics, Phys. Rev. Lett. **125**, 037203 (2020).

[2] W. Sun, W. Wang, H. Li, G. Zhang, D. Chen, J. Wang, and Z. Cheng, Controlling bimerons as skyrmion analogues by ferroelectric polarization in 2D van der Waals multiferroic heterostructures, Nat. Commun. **11**, 5930 (2020).

[3] T. d. C. S. C. Gomes, Y. Sassi, D. Sanz-Hernandez, S. Krishnia, S. Collin, M.-B. Martin, P. Seneor, V. Cros, J. Grollier, and N. Reyren, Neuromorphic weighted sums with magnetic skyrmions, Nat. Electron. **8**, 204 (2025).

[4] L. Zhou, M. Tanwani, P. Tong, P. Gupta, Y. Wu, Y. Huang, H. Tian, S. Das, and Z. Hong, Harness of room-temperature polar skyrmion bag in oxide superlattice, Nat. Commun. **16**, 9911 (2025).

[5] X. Wang, Z. He, Y. Dai, B. Huang, and Y. Ma, Magnetoelectric Bimeron in 2D Hexagonal Lattice, Adv. Funct. Mater. **36**, 202510581 (2026).

[6] N. Paul, Y. Zhang, and L. Fu, Giant proximity exchange and flat Chern band in 2D magnet-semiconductor heterostructures, Sci. Adv. **9**, eabn1401 (2023).

[7] E. Mascot, J. Bedow, M. Graham, S. Rachel, and D. K. Morr, Topological superconductivity in skyrmion lattices, npj Quantum Mater. **6**, 6 (2021).

[8] T. Hirosawa, J. Klinovaja, D. Loss, and S. A. Diaz, Laser-Controlled Real- and Reciprocal-Space Topology in Multiferroic Insulators, Phys. Rev. Lett. **128**, 037201 (2022).

[9] H.-Y. Chen, T. Nomoto, M. Hirschberger, and R. Arita, Topological Hall Effect of Skyrmions from first Principles, Phys. Rev. X **15**, 011054 (2025).

[10] O. K. Forslund, X. Liu, S. Shin, C. Lin, M. Horio, Q. Wang, K. Kramer, S. Mukherjee, T. Kim, C. Cacho *et al.*, Anomalous Hall Effect due to Magnetic Fluctuations in a Ferromagnetic Weyl Semimetal, Phys. Rev. Lett. **134**, 126602 (2025).

[11] Y. Wang, H. Sun, C. Wu, W. Zhang, S.-D. Guo, Y. She, and P. Li, Multifield tunable valley splitting and anomalous valley Hall effect in two-dimensional antiferromagnetic MnBr, Phys. Rev. B **111**, 085432 (2025).

[12] Z. Zhang, Y. Bai, X. Zou, B. Huang, Y. Dai, and C. Niu, Altermagnetic quantum spin Hall effect in a Chern homobilayer, Phys. Rev. B **112**, 085128 (2025).

[13] Y. Yu, X. Chen, X. Liu, J. Li, B. Sanyal, X. Kong, F. M. Peeters, and L. Li, Ferromagnetism with in-plane magnetization, Dirac spin-gapless semiconducting properties, and tunable topological states in two-dimensional rare-earth metal dinitrides, Phys. Rev. B **105**, 024407 (2022).

[14] Y. Yu, X. Xie, X. Liu, J. Li, F. M. Peeters, and L. Li, Two-dimensional semimetal states in transition metal

trichlorides: A first-principles study, Appl. Phys. Lett. **121**, 112405 (2022).

[15] W. Sun, C. Yang, W. Wang, Y. Liu, X. Wang, S. Huang, and Z. Cheng, Proposing Altermagnetic-Ferroelectric Type-III Multiferroics with Robust Magnetoelectric Coupling, Adv. Mater. **37**, 2502575 (2025).

[16] L. Bai, W. Feng, S. Liu, L. Smejkal, Y. Mokrousov, and Y. Yao, Altermagnetism: Exploring New Frontiers in Magnetism and Spintronics, Adv. Funct. Mater. **34**, 2409327 (2024).

[17] V. Leeb, A. Mook, L. Smejkal, and J. Knolle, Spontaneous Formation of Altermagnetism from Orbital Ordering, Phys. Rev. Lett. **132**, 236701 (2024).

[18] T. Jungwirth, J. Sinova, R. M. Fernandes, Q. Liu, H. Watanabe, S. Murakami, S. Nakatsuji, and L. Smejkal, Symmetry, microscopy and spectroscopy signatures of altermagnetism, Nature **649**, 837 (2026).

[19] C. Song, H. Bai, Z. Zhou, L. Han, H. Reichlova, J. H. Dil, J. Liu, X. Chen, and F. Pan, Altermagnets as a new class of functional materials, Nat. Rev. Mater. **10**, 473 (2025).

[20] A. K. C. Tan, P. Ho, J. Lourembam, L. Huang, H. K. Tan, C. J. O. Reichhardt, C. Reichhardt, and A. Soumyanarayanan, Visualizing the strongly reshaped skyrmion Hall effect in multilayer wire devices, Nat. Commun. **12**, 4252 (2021).

[21] Z. Jin, Z. Zeng, Y. Cao, and P. Yan, Skyrmion Hall Effect in Altermagnets, Phys. Rev. Lett. **133**, 196701 (2024).

[22] K. Dou, Z. He, J. Zhao, W. Du, Y. Dai, B. Huang, and Y. Ma, Engineering Topological Spin Hall Effect in 2D Multiferroic Material, Adv. Sci. **11**, 2407982 (2024).

[23] X. Li, J. Koo, Z. Zhu, K. Behnia, and B. Yan, Field-linear anomalous Hall effect and Berry curvature induced by spin chirality in the kagome antiferromagnet $Mn_3Sn$, Nat. Commun. **14**, 1642 (2023).

[24] P. Li, L. Tao, X. Jin, G. Wan, J. Zhang, Y.-F. Zhang, J.-T. Sun, J. Pan, and S. Du, Nonvolatile Multistate Manipulation of Topological Magnetism in Monolayer $CrI_3$ through Quadruple-Well Ferroelectric Materials, Nano Lett. **24**, 2345 (2024).

[25] L. Smejkal, J. Sinova, and T. Jungwirth, Emerging Research Landscape of Altermagnetism, Phys. Rev. X **12**, 040501 (2022).

[26] Y. Ga, Q. Cui, Y. Zhu, D. Yu, L. Wang, J. Liang, and H. Yang, Anisotropic Dzyaloshinskii-Moriya interaction protected by $D_{2d}$ crystal symmetry in two-dimensional ternary compounds, npj Comput. Mater. **8**, 128 (2022).

[27] C.-C. Wei, X. Li, S. Hatt, X. Huai, J. Liu, B. Singh, K.-M. Kim, R. M. Fernandes, P. Cardon, L. Zhao *et al.*, $La_2O_3Mn_2Se_2$: A correlated insulating layered *d*-wave altermagnet, Phys. Rev. Mater. **9**, 024402 (2025).

[28] X. Zhang, Y. Zhou, and M. Ezawa, Magnetic bilayer-skyrmions without skyrmion Hall effect, Nat. Commun. **7**, 10293 (2016).

[29] X. Zhang, G. Wan, J. Zhang, Y.-F. Zhang, J. Pan, and S. Du, Eliminating Skyrmion Hall Effect in Ferromagnetic

Skyrmions, Nano Lett. **24**, 10796 (2024).

[30] B. Pan, P. Zhou, P. Lyu, H. Xiao, X. Yang, and L. Sun, General Stacking Theory for Altermagnetism in Bilayer Systems, Phys. Rev. Lett. **133**, 166701 (2024).

[31] See Supplemental Material for detailed information on Methods, symmetry rules governing the interlayer DM vector, A-AFM stacking configurations under $\hat{P} = P T_z T_O$ symmetry and the associated system stability, strain-driven switching of topological and transport properties, and calculation details of magnetic parameters.

[32] T. Moriya, Anisotropic superexchange interaction and weak ferromagnetism, Phys. Rev. **120**, 91 (1960).

[33] Z. Liu, M. Ozeki, S. Asai, S. Itoh, and T. Masuda, Chiral Split Magnon in Altermagnetic MnTe, Phys. Rev. Lett. **133**, 156702 (2024).

[34] K. Wu, J. Dong, M. Zhu, F. Zheng, and J. Zhang, Magnon Splitting and Magnon Spin Transport in Altermagnets, Chin. Phys. Lett. **42**, 070702 (2025).

[35] W. Sun, H. Ye, L. Liang, N. Ding, S. Dong, and S.-S. Wang, Stacking-dependent ferroicity of a reversed bilayer: Altermagnetism or ferroelectricity, Phys. Rev. B **110**, 224418 (2024).

[36] W. Sun, W. Wang, C. Yang, S. Huang, N. Ding, S. Dong, and Z. Cheng, Designing Spin Symmetry for Altermagnetism with Strong Magnetoelectric Coupling, Adv. Sci. **12**, e03235 (2025).

[37] H.-Z. Liu, R. He, J.-Y. Zhan, D. Wang, M.-D. He, N. Luo, J. Zeng, K.-Q. Chen, and L.-M. Tang, Anomalous Hall effect in A-type antiferromagnetic bilayers, Phys. Rev. B **112**, 134411 (2025).

[38] Y. Che, H. Lv, X. Wu, and J. Yang, Bilayer Metal-Organic Framework Altermagnets with Electrically Tunable Spin-Split Valleys, J. Am. Chem. Soc. **147**, 14806 (2025).

[39] O. Chalaev and D. Loss, Spin-Hall conductivity due to Rashba spin-orbit interaction in disordered systems, Phys. Rev. B **71**, 245318 (2005).

[40] G. Y. Guo, S. Murakami, T.-W. Chen, and N. Nagaosa, Intrinsic spin Hall effect in platinum: First-principles calculations, Phys. Rev. Lett. **100**, 096401 (2008).

[41] J. H. Ryoo, C.-H. Park, and I. Souza, Computation of intrinsic spin Hall conductivities from first principles using maximally localized Wannier functions, Phys. Rev. B **99**, 235113 (2019).

[42] D. J. Thouless, M. Kohmoto, M. P. Nightingale, and M. den Nijs, Quantized Hall conductance in a two-dimensional periodic potential, Phys. Rev. Lett. **49**, 405 (1982).

[43] D. Xiao, M.-C. Chang, and Q. Niu, Berry phase effects on electronic properties, Rev. Mod. Phys. **82**, 1959 (2010).

[44] X. Tian, Z. Zhang, L. Guan, X. Liu, X. Zhao, and L. Li, Multiple magnetic states, valley electronics, and topological phase transitions in two-dimensional Janus *XYZ*H (*X* = Sc, Y, La; *Y* = Cl, Br, I; *Z* = S, Se, Te) monolayers and bilayers, Phys. Rev. B **112**, 035413 (2025).

[45] X. Liu, L. Guan, X. Liu, X. Kong, and L. Li, Two-dimensional quasilayered hexagonal $V_2N_2$ manipulated by temperature, strain, electric field, and magnetic field: Antiferromagnetic higher-order topological Mott state and in-plane valley polarization effect, Phys. Rev. B **113**, 054436 (2026).

[46] A. Thiele, Steady-state motion of magnetic domains, Phys. Rev. Lett. **30**, 230 (1973).

[47] W. Jiang, X. Zhang, G. Yu, W. Zhang, X. Wang, M. Benjamin Jungfleisch, J. E. Pearson, X. Cheng, O. Heinonen, and K. L. Wang, Direct observation of the skyrmion Hall effect, Nat. Phys. **13**, 162 (2017).

[48] X. Marti, I. Fina, C. Frontera, J. Liu, P. Wadley, Q. He, R. J. Paull, J. D. Clarkson, J. Kudrnovsky, I. Turek *et al.*, Room-temperature antiferromagnetic memory resistor, Nat. Mater. **13**, 367 (2014).

[49] B. Göbel, L. Schimpf, and I. Mertig, Topological orbital Hall effect caused by skyrmions and antiferromagnetic skyrmions, Commun. Phys. **8**, 17 (2025).

[50] W. Du, K. Dou, X. Li, Y. Dai, Z. Wang, B. Huang, and Y. Ma, Topological layer Hall effect in two-dimensional type-I multiferroic heterostructure, Nat. Commun. **16**, 6141 (2025).